\documentclass{article}
\usepackage{spconf}  
\usepackage[utf8]{inputenc}
\usepackage{amsmath}
\usepackage{amssymb}
\usepackage[hidelinks]{hyperref}
\usepackage{graphicx}
\usepackage{subcaption}
\usepackage{bm}

\title{Score-based Diffusion Models for Bayesian Image Reconstruction}
\name{Michael T. McCann$^1$, Hyungjin Chung$^2$, Jong Chul Ye$^2$, Marc L. Klasky$^1$
\thanks{\{mccann, mklasky\}@lanl.gov, \{hj.chung, jong.ye\}@kaist.ac.kr}
}
\address{
$^1$Los Alamos National Laboratory 
$^2$Korea Advanced Institute of Science \& Technology
}

\renewcommand{\vec}[1]{\bm{#1}}  

\renewcommand{\bar}{\mathop{|}}  

\newcommand{\x}{{\vec{x}}}  
\newcommand{\X}{{\vec{X}}}  

\newcommand{\y}{{\vec{y}}}  
\newcommand{\Y}{{\vec{Y}}}  

\newcommand{\normpdf}{{\mathcal{N}}}  

\newcommand{\fX}{f_{\X}}

\newcommand{\fXonY}{f_{\X \bar \Y}}
\newcommand{\fYonX}{f_{\Y \bar \X}}

\newcommand{\LMAP}{\mathcal{L}_\text{MAP}}

\newcommand{\xMAP}{\hat{\x}_\text{MAP}}

\newcommand{\xMMSE}{\hat{\x}_\text{MMSE}}
\newcommand{\sigmaMMSE}{\sigma_\text{MMSE}}

\newcommand{\z}{{\vec{z}}}  
\newcommand{\Z}{{\vec{Z}}}  
\newcommand{\fZ}{f_{\Z}}

\newcommand{\N}{{\vec{N}}}  

\newcommand{\recon}{{\hat{\x}}}  

\DeclareMathOperator{\E}{\mathbb{E}}
\DeclareMathOperator*{\argmin}{argmin}
\DeclareMathOperator*{\argmax}{argmax}

\usepackage{xcolor}

\begin{document}

\maketitle

\begin{abstract}
This paper explores the use of score-based diffusion models for Bayesian image reconstruction.
Diffusion models are an efficient tool for generative modeling.
Diffusion models can also be used
 for solving image reconstruction problems.
We present a simple and flexible algorithm
for training a diffusion model
and using it for maximum a posteriori reconstruction,
minimum mean square error reconstruction,
and posterior sampling.
We present experiments on both a linear and a nonlinear reconstruction problem
that highlight the strengths and limitations of the approach.

\end{abstract}
\begin{keywords}%
image reconstruction,
inverse problems in imaging,
Bayesian inference,
generative modeling,
diffusion models
\end{keywords}

\section{Introduction}
The goal of image reconstruction
(also called inverse problems in imaging)
is to
recover an image, $\x \in \mathbb{R}^N$,
from its measurements, $\y \in \mathbb{R}^M$.
Image reconstruction problems arise in a wide variety of fields.
For example,
the table of contents for the IEEE Transactions on Computational Imaging (2022, volume 8)
shows applications of image reconstruction algorithms in biomedical imaging, nondestructive testing, computational photography, remote sensing, astronomy, and more.
Approaches to image reconstruction have focused
on either using hand-designed priors
(e.g., in compressive sensing~\cite{donoho_compressed_2006,candes_robust_2006} and model-based image reconstruction~\cite{fessler_model_2010,beister_iterative_2012}),
learning simple models for images or their patches (e.g., dictionary learning~\cite{xu_low_2012,garcia-cardona_convolutional_2018}),
or using neural networks to learn priors~\cite{mccann_biomedical_2019,ongie_deep_2020},
usually in an \emph{implicit} way.
Recent advances in generative modeling---%
the process of learning a distribution of images from a training set
and generating novel samples from it---%
including
variational autoencoders~\cite{kingma_introduction_2019},
generative adversarial networks~\cite{arjovsky_wasserstein_2017a},
flows~\cite{kingma_glow_2018},
and diffusion models~\cite{ho_denoising_2020}
provide new opportunities to \emph{explicitly} learn complex priors
and use them for image reconstruction.

Here, we focus on the use of diffusion models for image reconstruction.
Diffusion models~\cite{ho_denoising_2020,song2021scorebased},
learn a data distribution by training a neural network to approximate its score function,
which is the gradient of the log density.
Most approaches involve using a sequence of noisy approximations
to the target density,
learning a score function at each noise level
(or, in the continuous case, as a function of the noise level).
These score functions can then be used to draw a new sample from
the target distribution
via an interactive process where the noise level is slowly decreased
 towards zero (sometimes called annealing).

Several works leverage diffusion models to solve image reconstruction problems when the measurement $\y$, and the corresponding likelihood $\fYonX$ is given.
However, due to intractability arising from multiple noise scales, some form of approximation is necessary.
One line of works~\cite{song2021scorebased,chung2022score,song2022solving,chung2022come} propose to alternate sampling steps with data consistency projections.
A more recent line of works~\cite{kawar2022denoising,chung2023diffusion} attempt to approximate the intractable likelihood to achieve a form of posterior sampling.

In this work,
we explore a simplified diffusion model scheme that uses only a single noise level.
This approach allows the model to be used in Bayesian image reconstruction, i.e.,
for maximum a posteriori reconstruction,
minimum mean square error reconstruction,
and posterior sampling,
without resorting to additional approximations.

\section{Background: Bayesian Reconstruction} \label{sec:image_recon}

\textbf{Maximum a posteriori (MAP) image reconstruction}
combines a model of the measurement process
with a model of the images to be reconstructed.
We model the measurements as realizations of a random variable $\Y$,
with PDF
$
\fYonX(\y \bar \x).
$
We model images as realizations of a random variable $\X$
with PDF
$
    \fX(\x).
$
This prior need not represent the actual distribution of images
(which may or may not even exist);
it is something that we design (or learn from data) to
represent our beliefs about what images we expect.
MAP seeks the image that is most likely conditional on the observed measurements:
$
    \xMAP (\y)= \argmax_\recon \fXonY(\recon \bar \y).
$
It can be convenient to use Bayes' theorem to express this as
\begin{equation} \label{eq:MAP}
     \xMAP (\y) \argmin_\recon -\log \fYonX(\y \bar \recon) - \log \fX(\recon).
\end{equation}

\textbf{Minimum mean square error (MMSE) image reconstruction}
seeks the image that will,
on average
conditional on the measurement,
be the closest to the ground truth:
\begin{equation} \label{eq:MMSE}
    \xMMSE(\y) = \argmin_\recon \E(\|\recon - \X \|^2_2 \bar \Y = \y).
\end{equation}
One can show that \eqref{eq:MMSE}
implies that
$
    \xMMSE(\y)  = \E(\X \bar \Y = \y),
$
and, further,
that
$
\E(\| \xMMSE(\Y) - \X \|^2_2)  \le \E(\| \recon(\Y) - \X \|^2_2)
$
for all  reconstruction algorithms $\recon$.
Therefore,
in the limit of an infinitely-large test set,
MMSE image reconstructor is
the algorithm that performs the best
(in terms of average MSE)
among all possible algorithms.
Put another way,
when papers compare methods in terms of signal-to-noise ratio (SNR)
or mean square error (MSE) on a testing set,
they are implicitly seeking an MMSE reconstruction algorithm.

Note that our statements about MMSE implicitly assume that the prior we design, $\fX$,
actually matches the distribution of the testing data,
i.e., the distribution over which the expectation is taken in the definition of the MMSE.

\section{Score-based data-driven priors}
In this section, 
we describe our proposed method to learn an (approximation of) $\fX$ directly from data
and use it to perform MAP reconstruction, MMSE reconstruction, and posterior sampling.
The key idea is that information about $\fX$ may be recovered from
an MMSE denoiser that acts on $\X$.

Assume that we can draw samples from the random variable $\X$,
which represents the distribution of images we aim to reconstruct
(or we have a sufficiently large training set of samples),
but that we do not know the PDF $\fX$, which we need to perform MAP and MMSE reconstruction.
We begin by forming a noisy version of $\X$, called $\Z$.
Let $\Z = \X + \N$,
with $\N \sim \normpdf(\vec{0}, \sigma^2 \vec{I})$,
for a value of $\sigma^2$ that we choose.
We then seek a function 
to solve the denoising problem,
i.e.,
to recover $\x$ from $\z$.
As already discussed,
the best possible denoiser (in an MSE sense) is the MMSE denoiser,
which is not known in general.
Any actual denoiser, e.g., a trained neural network,
only approximates MMSE denoising for a variety of reasons:
it is trained on a finite training set,
training finds only an approximate minimizer of the training loss,
and because any finite architecture cannot represent all possible functions.

Tweedie's formula~\cite{efron_tweedie_2011} relates the MMSE estimator to the PDF of the measurements.
It states that
\begin{equation}
    \xMMSE(\z) = \z + \sigma^2 \nabla \log \fZ(\z),
\end{equation}
where $\nabla \log \fZ(\z)$ is known as the \emph{score function}.
For a well-trained denoiser, we have $r(\z) \approx \xMMSE(\z)$
and therefore
\begin{equation} \label{eq:Tweedie}
    r(\z) \approx \z + \sigma^2 \nabla \log \fZ(\z).
\end{equation}

\subsection{Score-based MAP} \label{sec:score_MAP}
Recall that MAP reconstruction minimizes 
\begin{equation} \label{eq:score_MAP}
\LMAP(\x) =  -\log \fYonX(\y \bar \x) - \log \fX(\x).
\end{equation}
If we are willing to use $\fZ$ (the noisy version of the data distribution $\fX$)
as our prior,
we can use \eqref{eq:Tweedie} to write the gradient of the loss function as 
\begin{align} \label{eq:grad_score_MAP}
\nabla_\x \LMAP(\x) 
    &\approx  - \nabla_\x  \log \fYonX(\y \bar \x) -  \frac{1}{\sigma^2}(r(\x) - \x).
\end{align}
We can then use first order optimization methods to solve \eqref{eq:MAP}
and recover a MAP reconstruction.

\subsection{Posterior Sampling and Score-based MMSE}
The score function also provides a way to sample from the posterior, $\fXonY$,
which allows us both to estimate the MMSE reconstruction (which is the posterior mean)
and explore the uncertainty in the reconstruction.

To use the score function for sampling, we use
the theory of stochastic differential equations.
In particular,
the Euler–Maruyama discretization of overdamped Langevin Itô diffusion
(see \cite{li_stochastic_2019} equations (1), (2), and (4))
yields the iteration
\begin{equation} \label{eq:Langevin}
    \x^{k+1} = \x^{k} + \tau \nabla \log \fX(\x) + \sqrt{2\tau} \vec{E},
\end{equation}
where $\vec{E}$ is a draw from a standard multivariate normal distribution (zero mean, identity covariance)
and $\tau$ is a small positive step size.
We know that 
in the limit $k \to \infty$
and $\tau \to 0$
this iteration produces a distribution that approaches $\fX$
for any PDF $\fX$ for which \eqref{eq:Langevin} is well-defined,
i.e., the gradient exists.

Langevin diffusion \eqref{eq:Langevin} provides a way to draw samples from any distribution for which we know the score function.
This is useful both for qualitatively evaluating our prior $\fZ$
and for performing posterior sampling.
In the former case,
we substitute a well-trained denoiser for the score function,
resulting in the iteration
\begin{equation} \label{eq:score-based_sampling}
    \x^{k+1} = \x^{k} + \frac{\tau}{\sigma^2}(r(\x^k) - \x^k)  + \sqrt{2\tau} \vec{E}.
\end{equation}
Looking at the results of \eqref{eq:score-based_sampling}
provides a way to explicitly see what prior we have learned.

To sample the posterior,
we note that the score function for the posterior, $\fXonY$,
is exactly what we have already derived in Section~\ref{sec:score_MAP}.
Again taking $\fZ$ as our prior, we have
the iteration
\begin{multline} 
    \x^{k+1} = \x^{k} + \tau \nabla_\x  \log \fYonX(\y \bar \x^k) + \\ \frac{\tau}{\sigma^2}(r(\x^k) - \x^k) + \sqrt{2\tau} \vec{E}.
\end{multline}

\section{Experiments and Results}
We now describe our experiments and results,
including our training dataset,
design and training of neural networks,
sample generation,
and image reconstruction in a linear and nonlinear setting.

\subsection{Dataset and Neural Network Training}
We designed a synthetic dataset
to test our algorithms.
Our intent was to use a simple generative model
so that examples could be generate on-the-fly during network training,
but to include enough structure so as to learn a nontrivial prior.
We achieved these goals by generating IID Gaussian images,
filtering them repeatedly with a square averaging filter,
and thresholding the result.
This approach is efficient on a GPU and provides control over 
the smoothness of the resulting images.
For our experiments, we used $64 \times 64$ images
with 10 rounds of $3 \times 3$ averaging and a threshold set at 0.
We refer to this as the \emph{Gaussian blobs dataset}.

To learn a score function as described above, we trained neural network denoisers
on the Gaussian blobs dataset.
To explore the effect of the noise level,
we trained one network for a low noise level ($\sigma^2 = 10^{-2}$)
and one for a medium noise level ($\sigma^2 = 10^{-1}$).
Networks were convolutional neural networks (CNNs)
using the ReLU linearity,
a residual connection (i.e., $r(x) = r_\text{CNN}(x) + x$)
and with zero-padding at every layer.
We trained the networks using stochastic gradient descent 
using PyTorch~\cite{paszke_pytorch_2019}.
We performed a hyperparameter search over 
the learning rate ($10^{-\frac{4}{2}}$, $10^{-\frac{3}{2}}$, \dots,  $10^{\frac{0}{2}}$),
momentum (0, 0.9),
number of network layers (10, 15, 20),
number of network channels (16, 32, 64, 128),
and presence or absence of skip connections from the input to each hidden layer.
The batch size was 20.
Each of the 240 networks in the hyperparameter sweep was trained for 4 hours
on an NVIDIA GeForce RTX 2080 Ti.
We evaluated each network in terms of MSE loss on a validation set of 100 images.

In the low noise setting ($\sigma^2 = 10^{-2}$),
the best network had a validation loss of $2.06 \times 10^{-8}$
and trained for over 1.5 million steps.
It included skip connections,
had a depth of 15,
16 channels,
a learning rate of $10^{-\frac{1}{2}}$,
and momentum of 0.9.
Four other networks had validation losses below $10^{-7}$,
they differed in terms of depth (10, 15, 20)
and number of channels (16, 32, 64).

In the medium noise setting ($\sigma^2 = 10^{-1}$) ,
the best network had a validation loss of $4.06 \times 10^{-3}$
and trained for over 1.2 million steps.
It included skip connections,
had a depth of 10,
32 channels,
a learning rate of $10^{-\frac{3}{2}}$,
and momentum of 0.9.
Twelve other networks had validation losses below $4.2^{-3}$,
they differed in terms of depth (10, 15, 20),
number of channels (16, 32, 64, 128),
and learning rate ($10^{-\frac{3}{2}}$, $10^{-\frac{2}{2}}$,  $10^{-\frac{1}{2}}$).

These results suggest that 
skip connections and momentum promote
good denoising performance.
We did not see good performance from the same architecture
at two different learning rates,
which indicates that finding the correct learning rate for a given architecture is important.

\subsection{Score-based Sampling}
We used our trained denoisers in the context of score-based sampling
to evaluate qualitatively what distribution each had learned.
We ran the iteration \eqref{eq:score-based_sampling}
for 2,000 steps
with $\tau$ set such that $\tau / \sigma^2 = 0.1$.
We initialized the sampling with an IID Gaussian image
with mean 0.5 and variance equal to the $\sigma^2$ used during training.
To generate multiple samples,
we repeated this process with different random initializations.

\begin{figure}[h!]
    \centering
    \includegraphics[width=0.24\linewidth]{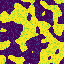}\hfill\hfill\hfill%
    \includegraphics[width=0.24\linewidth]{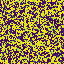}\hfill%
    \includegraphics[width=0.24\linewidth]{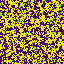}\hfill%
    \includegraphics[width=0.24\linewidth]{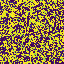}\\ \vspace{0.3em}
    \includegraphics[width=0.24\linewidth]{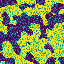}\hfill\hfill\hfill%
    \includegraphics[width=0.24\linewidth]{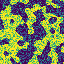}\hfill%
    \includegraphics[width=0.24\linewidth]{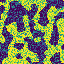}\hfill%
    \includegraphics[width=0.24\linewidth]{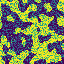}
    \caption{
    Left column: sample of target prior $\fZ$.
    Right columns: results of sampling from learned prior using \eqref{eq:score-based_sampling}.
    Top row, $\sigma^2 = 10^{-2}$; bottom row, $\sigma^2 = 10^{-1}$.
    }
    \label{fig:sampling}
\end{figure}

The results (Figure~\ref{fig:sampling}) show that the denoiser trained on the low noise level
does not produce reasonable samples of its target distribution,
while the denoiser trained at the moderate noise level does.
We believe this difference occurs because
the network trained at a moderate noise level 
must learn learn to combine information for neighboring pixels
to denoise well,
while the one trained at a low noise level 
acts mostly locally.
Based on these results, we use the model trained with $\sigma^2 = 10^{-1}$
in all subsequent experiments.

\subsection{Image Reconstruction}
We first evaluated our method on a \textbf{image inpainting problem}.
We drew a sample $\x$ from the Gaussian blobs dataset
and formed the measurement by drawing from $\Y = \vec{A}\x + \N_{\eta^2}$,
where $\vec{A}$ is a linear operator that crops away the center pixels of $\x$
(see Figure~\ref{fig:inpainting})
and  $\N_{\eta^2} \sim \normpdf(\vec{0}, \eta^2 \vec{I})$,
with $\eta^2 = 0.2$.
The gradient of the log likelihood was therefore
$
    \frac{1}{\eta^2}(\vec{A}^T\vec{A}\x - \vec{A}^T \y).
$
To perform MAP reconstruction, we solved the MAP objective \eqref{eq:score_MAP}
using gradient descent (500 steps, step size 0.1) with the gradient given by substituting this log likelihood into \eqref{eq:grad_score_MAP}.
To perform posterior sampling,
we substituted the log likelihood into \eqref{eq:score-based_sampling}
and ran 500 iterates with $\tau=10^{-3}$, restarting 500 times to generate 500 samples.
We formed the MMSE reconstruction by taking the pixelwise average of these samples.
We also computed the pixelwise standard deviation of the samples
as a measure of uncertainty of the MMSE reconstruction,
we we call $\sigma_\text{MMSE}$.

\newlength\panelwidth
\setlength\panelwidth{2cm}

\begin{figure}[htbp]
    \begin{subfigure}{.240\linewidth}
    \includegraphics[width=\panelwidth]{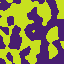} 
    \caption{$\x$}
    \end{subfigure}\hfill
    \begin{subfigure}{.240\linewidth}
    \includegraphics[width=\panelwidth]{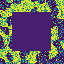} 
    \caption{$\y$}
    \end{subfigure}\hfill
    \begin{subfigure}{.480\linewidth}
    \includegraphics[width=\panelwidth]{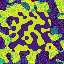}
    \includegraphics[width=\panelwidth]{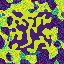} 
    \caption{MAP reconstructions, $\xMAP$}
    \end{subfigure}\\ 
    \begin{subfigure}{.480\linewidth}
    \includegraphics[width=\panelwidth]{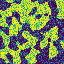}
    \includegraphics[width=\panelwidth]{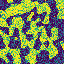}%
    \caption{posterior samples}
    \end{subfigure}\hfill
    \begin{subfigure}{.240\linewidth}
    \includegraphics[width=\panelwidth]{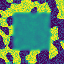} 
    \caption{$\xMMSE$}
    \end{subfigure}\hfill
    \begin{subfigure}{.240\linewidth}
    \includegraphics[width=\panelwidth]{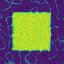} 
    \caption{$\sigmaMMSE$}
    \end{subfigure}     
    \caption{Inptainting results.}
    \label{fig:inpainting}
\end{figure}

Results for the inpainting problem are shown in Figure~\ref{fig:inpainting}.
We found that,
due to the nonconvexity of the MAP objective \eqref{eq:score_MAP},
the solution depended on the initialization.
By using different random initializations,
we could produce different MAP reconstructions.
These may have similar values of the MAP objective or not;
our framework does not allow us to check because we only have access to the gradient of the MAP objective,
with no way to compute its value.
Each MAP reconstruction reflected the properties of the forward model:
there is no information in the measurement about the center of the image $\x$,
therefore the center of the MAP reconstructions differ more than the edges.
On the edge,
the MAP reconstructions look noisy;
careful comparison between the measurement and reconstruction
reveals that this noise reflects the noise in the measurement---%
not the noisy prior;
this makes sense because the data fidelity term draws $\xMAP$ close to $\y$.
The posterior samples are also more similar to each other on the edge,
but they exhibit noise as a result of the noisy prior.
The MMSE reconstruction is constant in the center,
because MMSE reconstruction is equivalent to the expectation of the prior
when there is no information coming from the measurement.
The pixel-wise standard deviation of $\xMMSE$ reveals that most of the variability 
in the posterior samples is along the edges of blobs and in the center.

The PSNRs (with respect to the ground truth, peak = 1.0) of the posterior samples ranged from
4.72 dB to 6.29 dB;
the MAP PSNRs were 5.80 dB and 6.19 dB,
and the MMSE PSNR was 9.79.
The MAP data fidelities (PSNR of $\vec{A}\hat{\x}$ with respect to $\y$)
were 10.77 dB and 10.99 dB;
the MMSE data fidelity was 12.84 dB.
Note that the MMSE reconstruction is quantitatively the best
despite it looking qualitatively blurry.

We also solved a \textbf{nonlinear Fourier magnitude retrieval problem},
using the nonlinear operator $A(\x) = \mathcal{F}^{-1} (\mathcal{F}(\x)/|\mathcal{F}(\x)|)$,
where $\mathcal{F}$ denotes the Fourier transform and $| \cdot |$ is applied elementwise.
We set the measurement noise $\eta^2 = 10^{-4}$
and computed the log likelihood using automatic differentiation
(although it can also be derived by hand).
For MAP reconstruction, we used 5,000 steps with a step size of $10^{-3}$;
for posterior sampling, we used the same parameters as for the linear problem.

\begin{figure}[htbp]
    \begin{subfigure}{.240\linewidth}
    \includegraphics[width=\panelwidth]{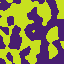} 
    \caption{$\x$}
    \end{subfigure}\hfill
    \begin{subfigure}{.240\linewidth}
    \includegraphics[width=\panelwidth]{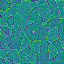} 
    \caption{$\y$}
    \end{subfigure}\hfill
    \begin{subfigure}{.480\linewidth}
    \includegraphics[width=\panelwidth]{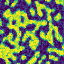}
    \includegraphics[width=\panelwidth]{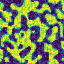} 
    \caption{MAP reconstructions, $\xMAP$}
    \end{subfigure}\\ 
    \begin{subfigure}{.480\linewidth}
    \includegraphics[width=\panelwidth]{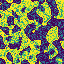}
    \includegraphics[width=\panelwidth]{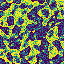}%
    \caption{posterior samples}
    \end{subfigure}\hfill
    \begin{subfigure}{.240\linewidth}
    \includegraphics[width=\panelwidth]{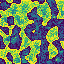} 
    \caption{$\xMMSE$}
    \end{subfigure}\hfill
    \begin{subfigure}{.240\linewidth}
    \includegraphics[width=\panelwidth]{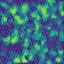} 
    \caption{$\sigmaMMSE$}
    \end{subfigure}     
    \caption{Magnitude retrieval results.}
    \label{fig:mag}
\end{figure}

Results for the magnitude retrieval problem are given in Figure~\ref{fig:mag}.
Here, the MAP reconstructions look smooth,
but they do not do a good job of capturing the edge locations in the data.
The posterior samples do capture the edges well,
and MMSE reconstruction is a good match to the ground truth.
The PSNRs of the posterior samples ranged from
-7.88 dB to 8.92 dB;
the MAP PSNRs were 4.82 dB and 5.14 dB,
and the MMSE PSNR was 11.40.
The MAP data fidelities
were 41.84 dB and 41.61 dB;
The MMSE data fidelity was 42.93 dB.
These results show that, while the MAP reconstructions do not visually resemble the ground truth,
the provide a high data fidelity,
meaning that they are reasonable solutions to the reconstruction problem.

\section{Discussion and Conclusions}
We have presented a method for using score-based diffusion models
in the context of Bayesian image reconstruction.
The advantage of our approach is that it is simple
and relies on only two approximations:
 the trained denoiser approximates the MMSE denoiser
and  the noisy prior $\fZ$ approximates the data prior $\fX$.
The first approximation is tightened in part by training the best possible denoiser,
but further work is needed to train denoisers that mimic the MMSE result on all possible inputs,
rather than just those that it is likely to see during training.
The second approximation can be made tighter by reducing the noise level during training;
however our work establishes that reducing this too much causes the model to fail to sample properly.
Future work will explore schemes for reducing this noise
while maintaining the ability to solve inverse problems 
in a straightforward way.
Our work also serves as a reminder that MMSE reconstructions,
while they maximize quantitative metrics such as MSE and SNR,
may not be qualitatively desirable,
e.g., consider the blurry MMSE result in Figure~\ref{fig:inpainting}.
This fact points towards using task-driven metrics 
such as segmentation or classification accuracy to evaluate the results of image reconstruction algorithms.

\bibliographystyle{IEEEbib}
\bibliography{refs_mike,refs_hyungjin}

\end{document}